\documentclass[aps,pre,twocolumn,showpacs]{revtex4}
\usepackage{epsf}

\newcommand{\Maybeeqs}{Equations~}
\newcommand{\maybeeq}{equation~}

\newcommand{\doubleintegral}{\int_0^{2\pi}\frac{dp_i}{2\pi}
\int_0^{2\pi}\frac{dp_j}{2\pi}}
\newcommand{\bcint}{\int_{3-\sqrt{8}}^1 dz}
\newcommand{\prob}{\mathrm{prob}}
\newcommand{\IdentityMatrix}{1}

\begin{document}

\title{Boundary conditions and defect lines in the Abelian
sandpile model}

\author{M. Jeng}

\email{mjeng@siue.edu}

\affiliation{
Box 1654, Department of Physics, Southern Illinois
University Edwardsville, Edwardsville, IL, 62025}

%-----------------------------------------------------------------%

\begin{abstract}

\noindent We add a defect line of dissipation, or crack, to
the Abelian sandpile model.  We find that the defect line 
renormalizes to separate the two-dimensional plane into 
two half planes with open boundary conditions.  We also 
show that varying the amount of dissipation at a boundary 
of the Abelian sandpile model does not affect the universality 
class of the boundary condition. We demonstrate that a
universal coefficient associated with height probabilities near the
defect can be used to classify boundary conditions.

\end{abstract}

\pacs{05.65.+b,45.70.-n}

\maketitle

%-----------------------------------------------------------------%

\section{INTRODUCTION}
\label{sec:intro}

The Abelian sandpile model (ASM) was introduced by Bak, 
Tang, and Wiesenfeld as a model of self-organized criticality~\cite{BTW}. 
This well-known model was designed to demonstrate how simple 
rules can drive a system to a critical point, and thus produce 
power laws, without any fine-tuning of parameters. It has thus 
been used to explain power laws in a wide range of 
systems---see~\cite{BakBook,SOCBook} for a review. Since the ASM was first
introduced, a number of variations on the model have been 
introduced---see~\cite{Variations} for a review. However, 
the original ASM still provides a simple, important, and 
robust model for the generation of power laws.

The ASM is defined on a square lattice. Each site $a$ of the
lattice has a height variable, $h_a$, which can be any
integer from 1 to 4, inclusive, where $h_a$ represents the number
of grains of sand at that site. At each
timestep, a grain of sand is added  to a random site of the
lattice. After the addition of the grain, any site 
with more than four grains of sand is unstable, and collapses,
losing four grains of sand, while each of its four neighbors
gains one grain of sand. Unstable sites are repeatedly
collapsed, until every site is stable---i.e. no site has more
than four grains. Then, the next timestep, another grain is
added, and the entire process is repeated.

The original ASM is spatially homogenous
(except for the boundaries, which break translational
invariance), and most modifications of the
sandpile model have kept this feature. 
However, here we consider the
effects of a crack, represented by a defect line, 
along which grains of sand can be lost; in other words,
along which the number of grains is not conserved. 
In previous studies, dissipation was added to the bulk of the 
ASM (not breaking translational invariance), and was shown
to take the ASM off its critical 
point~\cite{DissipBreaksSOC,dissip.1,dissip.2}.
Our defect line of dissipation
breaks translational invariance, and we show 
that it causes the two-dimensional plane of the ASM to renormalize 
into two half planes with open boundary conditions. This 
shows that cracks in the ASM are highly relevant, and
essentially cleave the sandpile into separate pieces.
We demonstrate this by looking at the universal coefficient
associated with the modification of unit height
probabilities at large distances from the defect, and at the
correlation function between unit height variables on 
opposite sides of the defect.
The Green function for an ASM with a defect line
is calculated in section~\ref{sec:Green.defect}, and
results for the height probabilities and correlations
are presented in section~\ref{sec:height.defect}.

For most models of interest in condensed matter physics, 
the bulk properties can be studied with the boundary playing 
little to no role. For example, the two-dimensional Ising 
Model is often studied on a torus, so as to eliminate boundary 
effects.  However, this is not possible for the ASM. In 
the bulk of the ASM, the number of grains of sand is conserved 
during each toppling. If this was true for all sites, then
eventually we would reach a state where topplings continued 
without end. The ASM thus needs sites with dissipation---that 
is, sites where the number of grains is not conserved.
The most natural way to do this is with open boundary
conditions; sites at the open boundaries become unstable
when they have more then four grains (just as in the bulk),
but have only three neighbors to send grains to,
and send the fourth grain ``off the edge,'' removing it 
from the system.  Since this dissipation is necessary for a
well-defined sandpile, the boundary plays a crucial role
in the ASM, even when we are focused at points in the bulk. 
Correlation functions far from the boundary are independent
of the boundary conditions, just as in other condensed
matter statistical mechanical models; but the presence of
dissipation somewhere in the ASM (e.g. at the open boundary)
is necessary for the model to be well-defined.

We consider the effects of varying amounts of dissipation
along a boundary, and show that any amount of dissipation at
the edge results in the open boundary universality class.
The Green function is calculated in 
section~\ref{sec:Green.boundary}, and results for the height probabilities
and correlations are presented in 
section~\ref{sec:height.boundary}.

These results are intuitively reasonable, since
dissipation should be relevant in regions of
the ASM where the particles have no other way to leave the
ASM. However, it was also possible that such modifications
could have resulted in new, as yet undiscovered,
boundary conditions or defect states.
For example, Bariev, and McCoy and Perk, added a 
line defect of modified bond strengths to the Ising Model, and 
found that they were able to continuously vary the
dimension of the spin operator along the defect by
varying the defect bond strength~\cite{Bariev,McCoy}.
This continual
variation occurred despite the fact that the Ising Model
only has three conformally invariant boundary conditions.

The ASM has been associated with a conformal field theory
(CFT). While CFT's are generally well understood, the
ASM is a logarithmic CFT (LCFT) (specifically, the $c=-2$
CFT), many aspects of which
are still not well understood~\cite{Flohr.Tehran.LCFT}. 
In particular, our
understanding of the boundary states of LCFT's is
still fragmentary, and recent results on 
boundaries of the $c=-2$ CFT have been partially 
contradictory~\cite{BDY.LCFT.0,BDY.LCFT.1,BDY.LCFT.2,BDY.LCFT.3,BDY.LCFT.4}.
Connections between the LCFT bounary states, and the
ASM boundary states were made in~\cite{Ruelle.ASM.BDY.LCFT},
but the ASM representation of the some of the $c=-2$
LCFT boundary states is still unknown. Modifications to
the ASM such as those described in this paper, and
searches for other boundary conditions, could
help eludicate these relationships.
Our results provide some evidence that
the open and closed boundary conditions are the only
possibilities for the ASM, although it is still possible
that further calculations in this vein could uncover
new boundary conditions.

The identification of boundary states as closed, or
open, or in some new, as yet undiscovered class, uses
arguments from CFT that the
coefficients associated with the falloff of expectation
values (height probabilities) at large distances from the
defect should be universal. Our results both use this
expected universality, and confirm it, since we find, for
example, that the coefficient is unaffected by varying a
free parameter corresponding to the amount of dissipation at
the boundary. This confirmation, while expected, is
valuable, given the anomalous and unsettled nature of 
boundary LCFT associated with the ASM. 
This is a particularly important
point in light of recent arguments that use this universality
to argue that the four height variables in the ASM must
correspond to different fields in the corresponding
CFT~\cite{Mine.use.universality}.

%-----------------------------------------------------------------%

\section{THE FORMALISM}
\label{sec:formalism}

Dhar pointed out ASM is highly analytically tractable because 
of its Abelian nature---the same state results whether grains 
of sand are added first at site $a$ and then at site $b$, 
or first at site $b$ and then at site $a$~\cite{DharFirst}.
This is the basis of a well-established formalism for analyzing 
the ASM---see~\cite{DharReview} for a review.  We only give 
a quick coverage of the essential points here. 

It is useful to first generalize the above description of 
the ASM, to allow for more complicated topplings. The dynamics 
of the model are described by a toppling matrix, $\Delta_{ab}$, 
where $a$ and $b$ label sites of the lattice. The 
dimension of $\Delta$ is equal to the number of sites in 
the lattice, so $\Delta$ becomes infinite-dimensional as the 
size of the lattice goes to infinity.
We say that site $a$ is unstable if its height $h_a$ is 
greater than $\Delta_{aa}$.  If site $a$ is unstable, then 
every height changes by $h_b\to h_b-\Delta_{ab}$ (including 
at the site $b=a$).  We have 
the standard ASM, with open boundary 
conditions, if $\Delta_{ab}$ is 4 when $a=b$, -1 when $a$ 
and $b$ are nearest neighbors, and 0 otherwise.

Dhar showed that the states of any sandpile, given certain 
general conditions on the form of $\Delta$, are divided 
into transient states, which occur with probability zero 
after long amounts of time, and recurrent states, which 
all occur with equal probability after long amounts of time.  
The number of possible recurrent configurations is given 
by det($\Delta$)~\cite{DharFirst}.

Furthermore, Majumdar and Dhar 
also showed how to find the probability
for a site to have height one, and the joint probability for 
two sites to both have height one (as well as other, more complicated
probabilities)~\cite{Dhar.UnitCorrelations,Dhar.CFT}. 
The toppling matrix is modified by
removing specific bonds, and changing the toppling condition
at certain sites. For example, if we want to force
site $a$ to have height 1, we change the toppling
matrix so that $\Delta_{aa}=1$, and remove three of the 
bonds to neighboring sites (setting 
$\Delta_{ab}=\Delta_{ba}=0$ for 
those bonds). With this modified toppling matrix, $\Delta'$, 
site $a$ is now guaranteed to have height 1, and
det($\Delta'$) gives the number of recurrent configurations 
with $h_a=1$.  While $\Delta$ and $\Delta'$ are infinite-dimensional 
matrices (for an inifnite lattice), $B\equiv\Delta'-\Delta$ 
is 0 outside of a 4x4 submatrix.  So 
det($\Delta'$)/det($\Delta$)=det($\IdentityMatrix+B \Delta^{-1}$)
is an easily computable 4 by 4 matrix determinant, which 
gives the probability that, in a randomly chosen recurrent 
configuration, the site $a$ will have height 1.
The same process, with a different (8 by 8) matrix $B$,
can be used to find two-point correlations of height 1 variables.  

This process requires us to calculate the Green function
$G\equiv\Delta^{-1}$. The Green function has long been
known for the standard ASM, where $\Delta$ is simply the 
lattice Laplacian~\cite{Spitzer}. However, in the following 
sections we will be dealing with different toppling
conditions, and so will need to calculate the Green function 
for these new $\Delta$'s.

%-----------------------------------------------------------------%

\section{GREEN FUNCTION FOR THE DEFECT LINE}
\label{sec:Green.defect}

We introduce a defect line (or crack) through the middle of
the ASM, allowing dissipation to take place along the
defect, and not just along the open boundary conditions.
We take the lattice to be size $M\times(2L-1)$,
with the x-dimension taking on the values 
$i=0,1,\dots, (M-1)$, and the y-dimension taking on the
values $j=-(L-1),-(L-2),\dots,(L-2),(L-1)$. We take open
boundary conditions along all edges, and put the
defect along the line $j=0$. Along this line, the height 
variable can take on the values $1,2,\dots,(4+m)$, where 
$m>0$. A site along the defect topples if its height is 
greater than $(4+m)$. When it topples, it sends one grain 
to each of its four neighbors, and $m$ grains of sand are 
dissipated (i.e. disappear from the sandpile).

When $m$ is a positive integer, the theory has its most
obvious physical interpretation, but 
the theory can be modified to give a sensible
interpretation for any rational, positive, value of 
$m$~\cite{DissipBreaksSOC}. 
If in each toppling, $c_1$ grains are toppled, and $c_2$
grains sent to each neighbor, where $c_1$ and
$c_2$ are integers, then the ratio of grains
dissipated to grains moved, 
$m/4 \leftrightarrow (c_1/(4c_2))-1$,
can be any rational integer.

The toppling matrix $\Delta$ is the same as for the
standard ASM, except that $\Delta_{aa}=4+m$ for sites $a$
along the defect. When $m=0$ it becomes the standard ASM.
The toppling matrix can be written as

\begin{eqnarray}
% Line 1
\label{eq:factorizedform}
\Delta_{(i,j),(i',j')} & = &
\delta_{ii'}\Delta^{(2)}_{jj'}+\delta_{jj'}\Delta^{(1)}_{ii'}
\ ,\mathrm{with}\\
% Line 2
\label{eq:bulk1DG}
\Delta^{(1)}_{ii'} & \equiv &
\left\{
\begin{array}{cl} 
2 & \mathrm{if}\  i=i' \\ 
-1 & \mathrm{if}\ i=i'\pm 1 \\ 
0 & \mathrm{otherwise} 
\end{array} \right.
\ \ ,\mathrm{and} \\
% Line 3
\Delta^{(2)}_{jj'} & \equiv &
\left\{\begin{array}{cl} 
2 & \rm{if}\  j=j'\neq 0 \\ 
m+2 & \rm{if}\  j=j'=0 \\ 
-1 & \rm{if}\ j=j'\pm 1 \\ 
0 & \rm{otherwise} 
\end{array} \right.
\end{eqnarray}

Since $\Delta$ is Hermitian, if we find all of its normalized 
eigenvectors, we can easily invert it. Suppose that the
eigenvectors of $\Delta$ are
$e_{\vec{p},\vec{x}}$, with eigenvalues
$\lambda_{\vec{p}}$.
$\vec{p}$ and $\vec{x}$ are two-dimensional vectors and the
number of possible values of $\vec{p}$ is equal to the
dimension of $\Delta$, which is in turn equal to the number
of sites in the lattice. Then

\begin{eqnarray}
G_{\vec{x},\vec{y}} & = &
\Delta^{-1}_{\vec{x},\vec{y}} \\
& = & \sum_{\vec{p}} {1 \over \lambda_{\vec{p}}}
e_{\vec{p},\vec{x}} \ e_{\vec{p},\vec{y}}\ \ .
\end{eqnarray}

\noindent The form of $\Delta$ in
\maybeeq\ref{eq:factorizedform}
implies that the eigenvectors of $\Delta$
factorize into eigenvectors of
$\Delta^{(1)}$ and $\Delta^{(2)}$.

% TRIAL SHORT REGION

We thus want the eigenvectors of $\Delta^{(2)}$. (The eigenvectors 
of $\Delta^{(1)}$ are not only simpler, but
immediately follow
from the eigenvectors of $\Delta^{(2)}$, by setting 
$m=0$.) $j$ and $j'$ range from $-(L-1)$ to $(L-1)$, so
$\Delta^{(2)}$ has $2L-1$ eigenvectors.  The eigenvectors 
fall in three classes. There are $(L-1)$ oscillatory eigenvectors 
that are antisymmetric about $j=0$, and have momenta $p$
evenly spaced between 0 and $\pi$ ($p=n\pi/L$,
$n\in \mathcal{Z}$, $1\leq n\leq (L-1)$). There are another
$(L-1)$ oscillatory eigenvectors that are symmetric
about $j=0$, and have momenta p in the range $0<p<\pi$,
where the $p$ solve a transcendental
equation; in the limit $L\to\infty$ these momenta
$p$ also become equally spaced between 0 and $\pi$. Finally,
there is one exponentially decaying eigenvector,
symmetric about $j=0$.

Since $\Delta$ is Hermitian, we can
immediately obtain its inverse from these eigenvectors.
The sums over the two oscillatory sets of eigenvectors 
each produce integrals
in the limit $L\to\infty$, $M\to\infty$,
using the Euler-MacLaurin formula.
The last, exponentially decaying, eigenvector 
produces a single, discrete
contribution to the Green function. Writing the Green
function as a sum of the contributions 
from the three classes of eigenvectors gives

\begin{widetext}

\begin{eqnarray}
G(i,j,i',j') & = & \sum_{a=1}^3 G^{(a)}(i,j,i',j') \\
G^{(1)}(i,j,i',j') & = &
\frac{1}{2}G_0(i-i',j-j')-\frac{1}{2}G_0(i-i',j+j') \\
G^{(2)}(i,j,i',j') & = & 
\frac{1}{2}G_0(i-i',\mid j\mid-\mid j'\mid)+
\delta G^{(2)}(i-i',\mid j\mid+\mid j'\mid) \\
G^{(3)}(i,j,i',j') & = &
\frac{(-1)^{j+j'}}{2\sqrt{k^2-1}}
\frac{m}{\sqrt{m^2+4}}
\left( K-\sqrt{K^2-1} \right)^{\mid i-i'\mid}
\left(-\frac{m}{2} - {\sqrt{m^2+4} \over 2} 
\right)^{-(\mid j\mid+\mid j'\mid)},
\end{eqnarray}

\noindent where we have defined

\begin{equation}
K \equiv 2+\frac{1}{2}\sqrt{m^2+4} \ .
\end{equation}

\noindent $G_0$ is the well-known bulk Green
function~\cite{Spitzer}:

\begin{equation}
G_0(i,j) \equiv \doubleintegral 
\frac{\cos(p_ii)\cos(p_jj)-1}{4-2\cos p_i-2\cos p_j} \ .
\end{equation}

\noindent We have also defined

\begin{eqnarray}
\nonumber
\delta G^{(2)}(i,j) & \equiv & 
\left(1-\frac{m^2}{2}\right)G^{(2a)}(i,j)+
m\left(G^{(2a)}(i,j-1)-G^{(2a)}(i,j+1)\right) \\*
& & -\frac{1}{2}\left(G^{(2a)}(i,j-2)+G^{(2a)}(i,j+2)\right)+c_m
\\
\label{eq:definitionG2a}
G^{(2a)}(i,j) & \equiv & \doubleintegral
\frac{\cos(p_ii)\cos(p_jj)-1}{4-2\cos p_i-2\cos p_j}
\frac{1}{m^2+4\sin^2p_j} \\
c_m & \equiv & \doubleintegral
\frac{1}{2-\cos p_i-\cos p_j}
\frac{\sin^2 p_j}{m^2+4\sin^2 p_j}
\end{eqnarray}

\end{widetext}

%\noindent The integral for $c_m^{(1)}$ can be done analytically, but the
%result is not useful for our calculations. 

We want the behavior of $G(i,j,i,j')$
for $\mid j \mid + \mid j' \mid$ large. The expansion of the
bulk Green function $G_0(0,j)$ for large $j$ is
well-known~\cite{Spitzer}:

\begin{equation}
\label{eq:bulkGreen}
G_0(0,j) \longrightarrow
-\frac{1}{2\pi}\log (j)
-\frac{1}{\pi} \left(\frac{\gamma}{2}+\frac{3}{4}\log 2\right)+
\frac{1}{24\pi j^2}+\dots \ ,
%\mathcal{O}\left(\frac{1}{j^4}\right) \ ,
\end{equation}

\noindent where $\gamma=0.577\dots$ is the Euler-Mascheroni
constant. 
We also need the behavior of $\delta G^{(2a)}(0,j)$ for 
$j$ large. The integral over $p_j$ in \maybeeq\ref{eq:definitionG2a} 
can be done exactly, and making the substitution $z=e^{ip_i}$ 
gives a contour integral around the unit circle.  The integrand 
has two poles inside the unit circle, 
%at $z=\pm(\sqrt{m^2+4}-m)/2$, 
but these give contributions which either decay exponentially 
with $j$, or are independent of $j$, neither of which affects
our height calculations; so these contributions 
can be dropped.  The algebraic 
$j$-dependence comes from the branch cut in the integrand, 
running from $z=3-\sqrt{8}$ to $z=3+\sqrt{8}$, which gives

\begin{eqnarray}
G^{(2a)}(0,j) & \rightarrow &
\frac{1}{\pi} \mathcal{P} \bcint
\frac{z^j-1}{z-1} f(z) \\
\nonumber
f(z) & \equiv & 
\frac{1}{\sqrt{-z^2+6z-1}}
\left(\frac{z^2}{(z^2-1)^2-m^2z^2}\right), \\
\end{eqnarray}

\noindent where $\mathcal{P}$ indicates that we take the
principal part of the integral. We can use this to find the 
behavior of $G^{(2a)}(0,j)$ for large $j$, by
separating out the contributions from $z$ near 1, and 
expanding in a Laurent series in $j$. This then
gives the expansion of $\delta G^{(2)}(0,j)$:

%We write the integral as

% XXXX BREAK FOR TWOCOLUMN XXX
%
%\begin{eqnarray}
%\nonumber
%\hspace{-0.3in} \bcint \frac{z^j-1}{z-1}f(z) & = &
%f(1)\bcint \frac{z^j-1}{z-1} \\
%\nonumber
%& & +\bcint \frac{f(z)-f(1)}{z-1} z^j \\
%& & - \bcint \frac{f(z)-f(1)}{z-1}
%\end{eqnarray}

% THIS COMMENTED VERSION FORMATS BETTER IN
% ONECOLUMN/PREPRINT FORMAT
%
%\begin{eqnarray}
%\nonumber
%\hspace{-0.3in} \bcint \frac{z^j-1}{z-1}f(z) & = &
%f(1)\bcint \frac{z^j-1}{z-1} +
%\bcint \frac{f(z)-f(1)}{z-1} z^j\\
%& & - \bcint \frac{f(z)-f(1)}{z-1}
%\end{eqnarray}

%\noindent The first integral is straighforward, while the
%third integral is independent of $j$, and so can be dropped. 
%For large $j$, the second integral gets its dominant contribution 
%for $z$ near 1, and we can expand its integrand as a Taylor series 
%around $z=1$ to get a Laurent series in $j$.
%Putting everything together, we get the expansion of 
%$\delta G^{(2a)}(0,j)$ for large $j$, and thus, ultimately,
%the expansion of $\delta G^{(2)}(0,j)$:

\begin{equation}
\delta G^{(2)}(0,j) =
\frac{1}{4\pi}\log j + \frac{1}{m\pi j}
+\frac{m^2-96}{48\pi m^2j^2} +\dots,
%+\frac{m^2+48}{6\pi m^3y^3}
%+(\mathrm{terms}\ \mathrm{independent}\ \mathrm{of}\ j)+
%\mathcal{O}\left(\frac{1}{j^3}\right)
\end{equation}

\noindent dropping terms independent of $j$.

%-----------------------------------------------------------------%

\section{HEIGHT PROBABILITIES FOR THE DEFECT LINE}
\label{sec:height.defect}

Now that we have the Green function for the defect line, 
we can use it to calculate the height correlations, using
the methods outlined in section~\ref{sec:formalism}.
We find that the probability for a site a distance $j$ from
the defect to have height 1 is

\begin{equation}
\label{eq:height.decay}
\prob \left(h_{(i,j)}=1\right) = \frac{2(\pi-2)}{\pi^3}
\left(1+\frac{1}{4j^2}-\frac{1}{2mj^3}+\dots \right) \ .
%\mathcal{O}\left(\frac{1}{j^4}\right)\right) \ .
\end{equation}

The constant term, $2(\pi-2)/\pi^3$, is the bulk probability
for a site to have height one, first calculated 
in~\cite{Dhar.UnitCorrelations}. It was also found
in~\cite{Dhar.UnitCorrelations} that the correlation
function between two height one operators is

% XXXX BREAK FOR TWOCOLUMN XXX
%
\begin{eqnarray}
\nonumber
& & \prob \left( h_{(i_1,j_1)}=h_{(i_2,j_2)}=1 \right) = \\
& & \qquad = \left( \frac{2(\pi -2)}{\pi^3} \right)^2
\left( 1 - \frac{1}{2r^4} + \dots \right),
\mathrm{where} \\
& & r\equiv\sqrt{(i_1-i_2)^2+(j_1-j_2)^2} \ .
\end{eqnarray}

% THIS COMMENTED VERSION FORMATS BETTER IN
% ONECOLUMN/PREPRINT FORMAT
%
%\begin{eqnarray}
%\hspace{-0.2in} \prob \left( h_{(i_1,j_1)}=h_{(i_2,j_2)}=1 \right) & = &
%\left( \frac{2(\pi -2)}{\pi^3} \right)^2
%\left( 1 - \frac{1}{2r^4} + \dots \right),
%\mathrm{where} \\
%& & r\equiv\sqrt{(i_1-i_2)^2+(j_1-j_2)^2} \ .
%\end{eqnarray}

Thus, the height one operator is a dimension 2 operator.
Based on the identification of the ASM as a conformal field
theory~\cite{Dhar.CFT,Mahieu.Ruelle}, the coefficient of
$1/j^2$, in expectation values of dimension 2 operators a
distance $j$ from a boundary, is expected to be a universal 
number characteristic of the boundary 
condition~\cite{Cardy.BoundaryCoeff}.
And, in fact, this coefficient of +1/4 in 
\maybeeq\ref{eq:height.decay} is exactly the coefficient
seen for the height one probability at large distances from
an open boundary condition, as shown by Brankov, Ivashkevich, and 
Priezzhev~\cite{Bdy.Falloff}. This indicates that,
upon renormalization, the defect line becomes an open
boundary.

It is only sensible to talk about conventional boundary
conditions at the defect line, if the two half-planes on either
side of the defect have somehow been separated.
Evidence that the defect renormalizes to separate
the half-planes can be seen by looking at correlation
functions of points on opposite sides of the defect.
If, upon renormalizing the defect, the two 
sides of the defect were still ``connected,'' we would
expect that height variables on opposite sides would
still fall off as $1/r^4$, since the height one operator
has dimension 2. (Calculations of correlation functions along
boundaries by Ivashkevich have shown that the height one
operator also has dimension 2 along open 
boundaries~\cite{Bdy.Correlations}.)
However, we find that

\begin{widetext}

\begin{eqnarray}
\nonumber
\prob \left(h_{(i,j)}=1,h_{(i,-j)}=1\right) 
-\prob \left(h_{(i,j)}=1\right) 
\prob \left(h_{(i,-j)}=1\right) = \\*
= \left(\frac{2(\pi-2)}{\pi^3}\right)^2
\left(-\frac{1}{8m^2j^6}+
\mathcal{O}\left(\frac{1}{j^7}\right)\right)
\end{eqnarray}

\end{widetext}

While the height variable is a dimension two operator, its
correlations across the defect fall off as $1/r^6$.
The coefficient of the $1/r^4$ term renormalizes to zero,
and the $1/r^6$ term is non-universal,
depending continuously on $m$. We thus conclude that
the defect renormalizes to generate two separate half-planes
with open boundary conditions. 

This is physically reasonable. Adding dissipation throughout 
the bulk of the ASM is known to be relevant, driving the 
system off 
criticality~\cite{DissipBreaksSOC,dissip.1,dissip.2}.
More recently, adding dissipation in the bulk was identified 
with adding the integral of a dimension 0 variable 
(the logarithmic partner of the identity) throughout the 
bulk~\cite{Mahieu.Ruelle}.  
It would thus appear that the local addition of dissipation
should be represented by a dimension zero operator, which
would mean that adding dissipation along a defect 
line should be relevant, as we have found here.  In short, 
cracks in the ASM cleave the plane into disconnected regions
with open boundary conditions.
Similar separation with relevant perturbations along a defect 
occurs in other models---see, for example~\cite{Ising.Defect}.  
However, these results were not inevitable; as already
noted, line defects added by Bariev, and McCoy and Perk,
to the Ising Model, resulted in a continual range of
defect lines, along which the dimension of the spin operator
could be continuously varied, despite the fact that the 
Ising Model only has three possible boundary conditions.

%-----------------------------------------------------------------%

\section{GREEN FUNCTION FOR THE MODIFIED BOUNDARY
CONDITIONS}
\label{sec:Green.boundary}

The identification of a line of dissipation with open
boundary conditions brings up the question of whether
any other universality classes of boundary conditions with 
dissipation are even possible. The open and closed boundary
conditions are the most natural to impose on the ASM,
but other boundary conditions than these two conventional
ones can be written down. We create new boundary conditions by
varying the amount of dissipation along the boundary, and
show that, regardless of the amount of dissipation, we
stay in the open boundary universality class (so long
as the amount of dissipation is nonzero---that is, 
so long as the boundary is not closed).  

It is convenient to change the dimensions of the lattice from
those of section~\ref{sec:Green.defect}.
We take the lattice to be of size $M\times L$,
with the x-dimension taking on the values 
$i=0,1,\dots, (M-1)$, and the y-dimension taking on the
values $j=0,1,\dots, (L-1)$.  We impose a modified
boundary condition along $j=0$, and open boundary conditions on
the other three edges. (In the end, we take the limits 
$L\to\infty$ and $M\to\infty$, so our results should anyway
be insensitive to the boundary conditions on these three
edges.)

We allow the height variable on the boundary, $j=0$, to 
take on values $1, 2,\dots,b$. Sites on the boundary
become unstable when their height is greater than b, at
which point they topple, giving one grain to each of their 
three neighbors, and dropping $b-3$ grains off the edge.  
For $b=3$ this is the closed boundary condition, and
for $b=4$ this is the open boundary condition.
For $b<3$, sand is generated with each toppling, rather than 
dissipated, creating the possibility of never-ending cycles 
of toppling. We
therefore only consider $b>3$. The system can be given a sensible 
interpretation for any rational value of 
$b\geq 3$~\cite{DissipBreaksSOC}.
The toppling matrix between sites $(i,j)$ and
$(i',j')$ can be written

\begin{equation}
\Delta_{(i,j),(i',j')} = 
\delta_{ii'}\Delta^{(3)}_{jj'}+\delta_{jj'}\Delta^{(1)}_{ii'}
\ ,
\end{equation}

\noindent where $\Delta^{(1)}$ was defined in
\maybeeq\ref{eq:bulk1DG}, and 

\begin{equation}
\Delta^{(3)}_{jj'} \equiv
\left\{\begin{array}{cl} 
2 & \rm{if}\  j=j'\neq 0 \\ 
b-2 & \rm{if}\  j=j'=0 \\ 
-1 & \rm{if}\ j=j'\pm 1 \\ 
0 & \rm{otherwise} 
\end{array} \right. \\
\end{equation}

As with the defect, if we can find all the eigenvectors of
$\Delta$, we can easily invert it. 
$\Delta^{(3)}$, being $L$-dimensional, has $L$
eigenvectors.
When $3<b<5$,
$\Delta^{(3)}$ has $L$ eigenvectors that are oscillatory
functions of $j$, with momenta $p$, which satisfy a
transcendental equation. 
%(similar to 
%\maybeeq(\ref{eq:TranscendentalDefect})). 
In the
limit $L\to\infty$, the momenta are evenly spaced over
the range $0<p<\pi$. When $b>5$, $\Delta^{(3)}$ only has
$(L-1)$ such oscillatory eigenvectors, and one last 
eigenvector that is exponentially decaying in $j$.

%It is not hard to check
%that the eigenvectors of $\Delta^{(3)}_{jj'}$ are
%
%\begin{eqnarray}
%\nonumber
%e_p^j & = & \sqrt{\frac{2}{L}}
%\frac{1}{\sqrt{(b-4)^2+1+2(b-4)\cos(p)}} \\
%& & \qquad\qquad  \times \left((b-4)\sin(pj)+\sin(p(j+1))\right) \\
%\lambda_p & = & 2-2\cos p  \ ,
%\end{eqnarray}
%
%\noindent where $p$ is in the range $0<p<\pi$, and satisfies
%
%\begin{equation}
%\label{eq:TranscendentalBdy}
%\tan (Lp)=\frac{\sin p}{4-b-\cos p} \ .
%\end{equation}
%
%These eigenvectors are normalized in the limit $L\to\infty$.
%When $3<b<5$, \maybeeq\ref{eq:TranscendentalBdy} 
%has exactly one solution in each range $n\pi/L < p < (n+1)\pi/L$, 
%for every integer $n=0,1,2,\dots,(L-1)$.
%So \maybeeq\ref{eq:TranscendentalBdy} 
%provides all $L$ eigenvectors of $\Delta^{(3)}$ for $3<b<5$.  When 
%$b>5$, \maybeeq\ref{eq:TranscendentalBdy} has no solution 
%in the range $(L-1)\pi/L < p < \pi$ (in the limit
%$L\to\infty$), so that \maybeeq\ref{eq:TranscendentalBdy}
%provides only $(L-1)$ of the $L$ eigenvectors of
%$\Delta^{(3)}$. For $b>5$
%there is a solution with imaginary $p$; in the limit 
%$L\to\infty$, this last eigenvector is
%
%\begin{eqnarray}
%\label{eq:SpecialEvec}
%e^j & = & \sqrt{(b-4)^2-1} \ (4-b)^{-j-1} \\
%\mathrm{with}\ \mathrm{eigenvalue}\\
%\lambda & = & \frac{(b-3)^2}{b-4} \ .
%\end{eqnarray}

In the Green function, the summation 
over oscillatory eigenvectors can be turned into an integral in the 
limit $L\to\infty$, $M\to\infty$, with the Euler-MacLaurin 
formula. For $b>5$, the single, exponentially decaying
eigenvector produces a separate, discrete contribution to
the Green function. The Green function is then given by

\begin{widetext}

% THE HSPACE OF -0.7 INCHES WORKS WELL FOR THE PREPRINT
% CLASS. IN THE TWOCOLUMN FORMAT, IT WOULD BE BETTER TO
% HAVE NO HSPACE COMMAND.
\begin{eqnarray}
\label{eq:Green.Mod.Bdy.1}
G(i,j,i',j') & = & \tilde{G}(i,j,i',j') +
\theta(b-5)G^{\rm exp}(i,j,i',j') \quad, \mathrm{where} \\
\nonumber
\label{eq:tildeG.def}
\tilde{G}(i,j,i',j') & \equiv & 
\doubleintegral \frac{\cos (p_i(i-i'))}{2-\cos p_i-\cos p_j}
\frac{1}{(b-4)^2+1+2(b-4)\cos p_j} \\*
& & \hspace{-0.7in} \times \left[ (b-4)\sin(p_jj)+\sin(p_j(j+1))\right]
\left[ (b-4)\sin(p_jj')+\sin(p_j(j'+1))\right] \
,\mathrm{and} \\
G^{exp}(i,j,i',j') & \equiv &
\frac{(b-3)(b-5)}{2\sqrt{k^2-1}}(4-b)^{-j-j'-2}
(k-\sqrt{k^2-1})^{\mid i-i'\mid} \ .
\label{eq:Green.Mod.Bdy.4}
\end{eqnarray}

\end{widetext}

\noindent We have defined

\begin{eqnarray}
k & \equiv & 1+\frac{(b-3)^2}{2(b-4)} \ \ ,\mathrm{and} \\
\theta (x) & \equiv & 
\left\{ 
\begin{array}{cc} 
1 & \mathrm{if}\ x>0 \\
0 & \mathrm{if}\ x\leq 0
\end{array}
\right.
\end{eqnarray}

At first sight, this equation for the Green function seems 
to indicate that G has a slope discontinuity at $b=5$. However, 
this is not the case. $\tilde{G}$ is not smooth at $b=5$, 
and expanding $\tilde{G}$ as a function of $b$ near $b=5$ 
shows that the combination $\tilde{G}+\theta(b-5)G^{\rm 
exp}$ is actually smooth to all powers of $(b-5)$. Inspection 
shows that $G$ is also a smooth function of $b$ for all 
other $b$ in the range $3<b<\infty$ (including $b=4$).  

We need the Green function in two limits. First, for
$i=i'$ and $j+j'$ large, it is useful to write

\begin{widetext}

\begin{eqnarray}
\tilde{G}(i,j,i',j') & = & G_0(i-i',j-j') - G_0(i-i',j+j') +
\delta G(i-i',j+j') \\
\nonumber
\delta G(i,j) & \equiv & \doubleintegral
\frac{\cos (p_ii)}{2-\cos p_i-\cos p_j}
\frac{\sin p_j}{(b-4)^2+1+2(b-4)\cos p_j} \\
& & \times \left[\sin p_j \cos(p_jj)+(b-4+\cos p_j)\sin(p_jj)\right] 
\ .
\end{eqnarray}

\end{widetext}

\noindent In $\delta G(0,j)$, we can do the integral over
$p_i$ exactly, and then set $z=e^{ip_j}$. 
As before, the main contribution comes from the branch cut
between $z=3-\sqrt{8}$ and $z=1$. Expanding the integral
near $z=1$ gives

%\begin{equation}
%\delta G(0,j) = \oint \frac{dz}{2\pi i} 
%\frac{1}{\sqrt{z^2-6z+1}} \frac{(1+z)z^j}{1+(b-4)z}\ \ ,
%\end{equation}
%
%\noindent where $z$ is integrated over the unit circle in the
%complex plane. If $3<b<5$, this has no poles in the unit
%circle, and if $b>5$, it has a pole, but the pole gives a 
%contribution which decays exponentially with $j$. The
%algebraic contribution comes from the branch cut
%between $z=3-\sqrt{8}$ and $z=3+\sqrt{8}$. Expanding the
%integral near $z=1$ then gives

\begin{equation}
\delta G(0,j) \approx \frac{1}{\pi (b-3) j}
-\frac{1}{\pi (b-3)^2 j^2} +
\frac{b^2-8b+19}{2\pi (b-3)^3 j^3} +\dots
\end{equation}

We also need the expansion of the Green function along the
defect---that is, for $j=j'=0$ and $\mid i-i'\mid \gg 0$.
Using similar methods as before, we find

%We will also need the expansion of the Green function along the
%defect---that is, for $j=j'=0$ and $\mid i-i'\mid \gg 0$.
%Returning to \maybeeq\ref{eq:tildeG.def}, when $j=j'=0$
%we can do the $p_j$ integral exactly, and then
%substitute $z=e^{ip_i}$ to get a contour integral around 
%the unit circle. The integral has no poles inside the
%unit circle, and can be deformed to an integral over the
%branch cut from $z=3-\sqrt{8}$ to $z=3+\sqrt{8}$. Expanding
%the integral near $z=1$ gives

% XXXX BREAK FOR TWOCOLUMN XXX
%
\begin{eqnarray}
\nonumber
\tilde G(x=\mid i-i'\mid,j=j'=0) & \approx & \\*
& & \hspace{-1.45in} \approx \frac{1}{\pi (b-3)^2x^2}
-\frac{b^2-18b+57}{2\pi (b-4)^4x^4}+
\mathcal{O}\left(\frac{1}{x^6}\right)
\end{eqnarray}

% THIS COMMENTED VERSION FORMATS BETTER IN
% ONECOLUMN/PREPRINT FORMAT
%
%\begin{equation}
%\tilde G(x=\mid i-i'\mid,j=j'=0) \approx \frac{1}{\pi (b-3)^2x^2}
%-\frac{b^2-18b+57}{2\pi (b-4)^4x^4}+
%\mathcal{O}\left(\frac{1}{x^6}\right) \ .
%\end{equation}

%-----------------------------------------------------------------%

\section{HEIGHT PROBABILITIES FOR MODIFIED BOUNDARY
CONDITIONS}
\label{sec:height.boundary}

Using the Green function for modified boundary conditions,
we can calculate unit height probabilities with
the methods described in section~\ref{sec:formalism}.
We find that the probability for a site a distance $j$ from
the boundary to have height 1 is

\begin{equation}
\prob \left(h_{(i,j)}=1\right) \hspace{-0.025in} = \hspace{-0.025in} 
\frac{2(\pi-2)}{\pi^3}
\left(1+\frac{1}{4j^2}-\frac{1}{2(b-3)}\frac{1}{j^3}+
\dots\right)
%\mathcal{O}\left(\frac{1}{j^4}\right)\right) \ .
\end{equation}

As discussed earlier, the coefficient of the 
$1/j^2$ term is expected to be a universal number
characteristic of the boundary
condition~\cite{Cardy.BoundaryCoeff}, and is equal to
$+1/4$ for the open boundary condition~\cite{Bdy.Falloff}. 
We see here that the coefficient is  $+1/4$, and
independent of $b$ for 
$b>3$.  This both confirms the expectation 
that the coefficient should be universal, and indicates 
that the boundary is in the open boundary universality class
for any amount of dissipation ($b>3$). 

Note that the coefficient of $1/j^3$ is
non-universal, and diverges as $b\to 3$, indicating that
$b=3$ is a special point as we vary $b$. $b=3$
corresponds to the closed boundary condition, and
it is already known that the coefficient of the
$1/j^2$ term is different (-1/4) for the closed
boundary conditions; this is appropriate, since the closed
and open boundary conditions are clearly in different
universality classes~\cite{Bdy.Falloff}.

Boundary correlations along the $j=0$ boundary can
be calculated, and contain no surprises. The
correlation function between sites $(i,0)$ and 
$(i',0)$ falls off as 
\hfil\break $1/\mid i-i'\mid^4$ for
all values of $b$:

\begin{widetext}

% THE HSPACE OF -4.0 INCHES WORKS WELL FOR THE PREPRINT
% CLASS. -2.0 INCHES WORKS BETTER FOR THE TWOCOLUMN FORMAT.
\begin{eqnarray}
\nonumber
\prob \left(h_{(i,0)}=1,h_{(i',0)}=1\right) 
-\prob \left(h_{(i,0)}=1\right) 
\prob \left(h_{(i',0)}=1\right) & = & \\*
& & \hspace{-4.0in} = -
\left(\frac{(1-c_0+c_2)(-1+3c_0-4c_1+c_2)}{(b-3)\pi}\right)^2
\frac{1}{\mid i-i'\mid^4}+
\mathcal{O}\left(\frac{1}{\mid i-i'\mid^6}\right)
\label{eq:coeff}
\end{eqnarray}

\end{widetext}

\begin{figure}[tb]
\begin{center}
\epsfbox{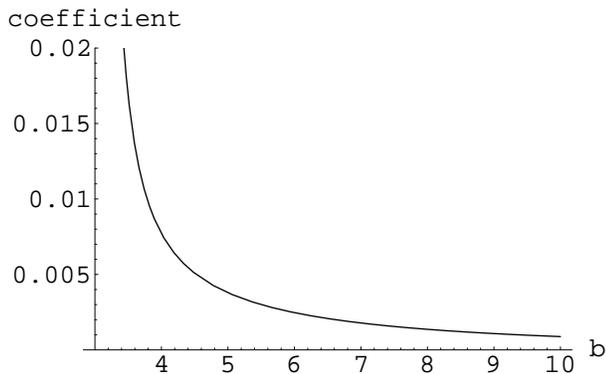}
\end{center}\caption{Coefficient of the $1/r^4$ term in the
two-point correlation function}
\label{fig:coeff}
\end{figure}

\noindent We have defined
$c_x\equiv G(x=\mid i-i'\mid,j=j'=0)$.
\Maybeeqs\ref{eq:Green.Mod.Bdy.1}-\ref{eq:Green.Mod.Bdy.4}
can be used to find analytic expressions for
$c_x$, for $x=0,1,2$.
However, the expressions are long and 
not particularly enlightening, so are not presented here. 
The (absolute value of the) coefficient of 
the $1/\mid~i-i'\mid^4$ term is plotted in figure~\ref{fig:coeff}. 
It falls off smoothly with increasing 
$b$.

The coefficient of $1/\mid i-i'\mid^4$ in
\maybeeq\ref{eq:coeff} diverges as b approaches 3,
reflecting the fact that $b=3$ is a 
fixed point of the renormalization
group flows, leading to non-smooth behavior in
physical properties.
However, the Green function, and
height correlations calculated from it, are perfectly
smooth as we vary $b$ through 4. It would appear that the
RG flows take us from $b=3$ to $b=\infty$, and that
$b=4$ is not a fixed point of the RG flows. However,
$b=\infty$ is in a sense the same as $b=4$, in that
both equally well represent the open boundary condition;
if $b=\infty$, the sites $j=0$ can hold an infinite number
of grains, and never topple---the sandpile thus acts as
if $j=1$ was the boundary, with an open boundary condition,
where grains fall ``off the edge'' to $j=0$.

We have shown that the addition of dissipation along a
defect line separates the ASM into two half-planes, each
with open boundary conditions. This brought up the question
of whether there are other universality classes of 
boundary conditions, with
varying amounts of dissipation along the
boundary. We find that any
amount of dissipation along a boundary results in the open
boundary universality class at large distances.
Classes of boundary conditions were identified
by the universal coefficient of the unit height
probability, far from the boundary or defect.

\bigskip

\acknowledgments{This work was supported in part by a Southern Illinois
University Edwardsville Summer Research Fellowship.}

%-----------------------------------------------------------------%

\end{document}